\documentclass[showpacs, aps, prb, unsortedaddress,showkeys,longbibliography,superscriptaddress ]{revtex4-1}

\usepackage{amsmath,amssymb,amsthm}
\usepackage{epsfig}
\usepackage{arydshln}

\newcommand{\nonum}{\nonumber \\}
\newcommand\eq[1] {(\ref{#1})} 
\newcommand{\beqa}{\begin{eqnarray}}
\newcommand{\eeqa}[1]{\label{#1}\end{eqnarray}}
\newcommand{\beq}{\begin{equation}}
\newcommand{\eeq}[1]{\label{#1}\end{equation}}

\newcommand{\Md}{\partial}

\newcommand{\Gd}{\delta}

\newcommand{\Gve}{\varepsilon}

\newcommand{\Gg}{\gamma}

\newcommand{\Gm}{\mu}

\newcommand{\Go}{\omega}

\newcommand{\bpm}{\begin{pmatrix}}
\newcommand{\epm}{\end{pmatrix}}

\usepackage{url}
\usepackage{hyperref}

\usepackage{graphics}
\usepackage{amssymb}
\begin{document}
\newpage
\title{Further comments on Mark Stockman's article "Criterion for Negative Refraction with Low Optical Losses from a Fundamental Principle of Causality"}

\date{\today}
\author{Graeme W. Milton}
\thanks{Corresponding Author}
\affiliation{Depsrtment of Mathematics,
155 S 1400 E RM 233, Salt Lake City, Utah 84112 }
\email{milton@math.utah.edu}
\author{Ankit Srivastava}
\affiliation{Department of Mechanical, Materials, and Aerospace Engineering,
Illinois Institute of Technology, Chicago, IL, 60616
USA}
\email{asriva13@iit.edu}

\begin{abstract}
We clarify the claims and errors in the paper "Criterion for Negative Refraction with Low Optical Losses from a Fundamental Principle of Causality" by Mark Stockman. Contrary
to the central assertion in that paper, simple examples consistent with the basic inequality which Stockman discovered show that it is possible to have negative refraction and low loss in an arbitrarily large frequency window. Further examination of the paper reveals additional errors that invalidate his argument that active
materials cannot have low loss and negative refraction in a frequency window. Also, we point out that for active materials non-analyticity of the electrical permittivity in the upper half complex frequency plane does not necessarily imply noncausality, as Stockman infers. 

\end{abstract}


\maketitle

\section{}
\setcounter{equation}{0}

In the paper \cite{stockman2007criterion}, Stockman considers the relative electrical permittivity $\Gve(\Go)=\Gve'(\Go)+i\Gve''(\Go)$
and magnetic permeability $\Gm(\Go)=\Gm'(\Go)+i\Gm''(\Go)$ and
derives the interesting inequality 
\beq J(\Go_0)\equiv \frac{2}{\pi}\int_0^\infty \frac{\epsilon''(\Go_1)\mu'(\Go_1)+\epsilon'(\Go_1)\mu''(\Go_1)}{(\omega_1^2-\omega_0^2)^2}\omega_1^3d\omega_1\leq -1
\eeq{1.0}
that holds at any real frequency $\Go_0$ where one has negative refraction. We emphasize that
this inequality is based on the analyticity properties of $\Gve(\Go)$ and $\Gm(\Go)$.
Therefore any models of  $\Gve(\Go)$ and $\Gm(\Go)$ satisfying these analyticity properties
automatically satisfy \eq{1.0} at frequencies where the refraction is negative.  

On the basis of this inequality Stockman claimed that "this criterion imposes the lower limits on the electric and magnetic losses in the region of the negative refraction'',
which is the main conclusion of the paper. It is generally accepted that
Stockman's claim is incorrect. Shortly after the original paper, Mackay and Lakhtakia \cite{mackay2007comment} proposed a causal material model which exhibited negative refraction and low (but not zero) loss in a finite frequency range. Stockman noted \cite{stockman2007stockman} that their proposal was unphysical and, in fact, was not in disagreement with the essence of the original claim. Subsequently, however, various authors \cite{kinsler2008causality, gralak2010macroscopic, srivastava2020causality} considered much simpler and widely used material models which are also at odds with Stockman's claim. The double-plasmon resonance model and the well known Lorentzian model of permittivity and permeability both allow for the existence of finite frequency ranges over which negative refraction can exist with zero or small loss. We explicitly show here that the lossless Lorentzian model also satisfies Stockman's inequality, at least in representative examples. We also note here that negative refraction does not require the existence of negative effective properties. Negative refraction can appear even in extremely simple systems such as a laminate, with laminations perpendicular to the interface, through Bragg diffraction \cite{willis2015negative,Srivastava2016}. While in diffraction gratings it is well known that rays of different diffraction orders 
in transmission are not constrained to lie on one side of the normal, it is not obvious that one can get pure negative refraction, with no beams having positive refraction, as these papers show. Diffraction by these laminates were studied in the early 1980's \cite{botten1981dielectric,botten1981finitely}, but no specific mention of negative refraction seems to have been made.

While \eq{1.0} does provide an unexpected relation giving restrictions on the minimum loss, as
governed by $\Gve''(\Go)$ and $\Gm''(\Go)$, it does not exclude these losses being zero in a frequency interval, and in particular in any interval
where one has negative refraction.

Mathematically, there is nothing preventing  $\Gve(\Go)$ and $\Gm(\Go)$
both being lossless and negative in a frequency window. Throughout this frequency window, the refractive index
will be negative. This is clearly seen through two examples. The simplest is when one
follows Kinsler and McCall \cite{kinsler2008causality} and takes a double-plasmon
resonance model (simplified to be lossless):
\beq \Gve(\Go)=\Gm(\Go)=1-\Go_{p}^2/\Go^2, \eeq{1.10}
in which $\Go_{p}$ is the plasma frequency, that is real and positive. Both $\Gve(\Go)$ and $\Gm(\Go)$ are negative for all positive real $\Go$ less than $\Go_{p}$.
Arbitrarily large values of  $\Go_{p}$ are still consistent with the analyticity properties of $\Gve(\Go)$ and $\Gm(\Go)$,
and so mathematically there is nothing preventing negative refraction over an arbitrarily large frequency window.
One may argue that a plasmonic model of $\Gm(\Go)$ is unrealistic. So, for our second example we take the more realistic Lorentzian model where
\beq \Gve(\Go)=1-\frac{\Go_{ep}^2}{\Go^2+i\Go\Gg_e-f_e^2}, \quad \Gm(\Go)=1-\frac{\Go_{mp}^2}{\Go^2+i\Go\Gg_m-f_m^2},
\eeq{1.15}
and take the limit where the loss parameters  $\Gg_e$ and $\Gg_m$ approach zero, giving
\beq \Gve(\Go)=1-\frac{\Go_{ep}^2}{\Go^2-f_e^2}-\frac{i\pi \Go_{ep}^2}{2f_e}\left[\delta(\omega+f_e)-\delta(\omega-f_e)\right],
\eeq{1.16}

\beq \Gm(\Go)=1-\frac{\Go_{mp}^2}{\Go^2-f_m^2}-\frac{i\pi \Go_{mp}^2}{2f_m}\left[\delta(\omega+f_m)-\delta(\omega-f_m)\right].
\eeq{1.16b}
In this model $\Gve(\Go)$ is real and negative in the frequency interval $f_e<\Go<\sqrt{f_e^2+\Go_{ep}^2} $, and similarly $\Gm(\Go)$ is real and negative in the frequency interval $f_m<\Go< \sqrt{f_m^2+\Go_{mp}^2}$. These overlap over a finite frequency interval when the larger of $(f_e,f_m)$ is less than the smaller of $(\sqrt{f_e^2+\Go_{ep}^2},\sqrt{f_m^2+\Go_{mp}^2})$. This is the frequency region which corresponds to negative refraction. Gralak and Tip \cite{gralak2010macroscopic} noted that such Lorentzian models can give rise to values $\Gve(\Go_0)=\Gm(\Go_0)=-1$ at a single frequency $\Go_0$, and it is just as clear to see that $\Gve(\Go)$ and $\Gm(\Go)$ can both be negative over a frequency interval \cite{srivastava2020causality}.

Naturally, one still gets negative refraction if one adds a small amount of loss in the frequency windows: for the Lorentzian model this may be accomplished by simply letting $\Gg_e$ and $\Gg_m$ take small positive values, and a small loss can similarly be added to the double-plasmon resonance model \cite{kinsler2008causality}.

\begin{figure}[htp]
\includegraphics[scale=.45]{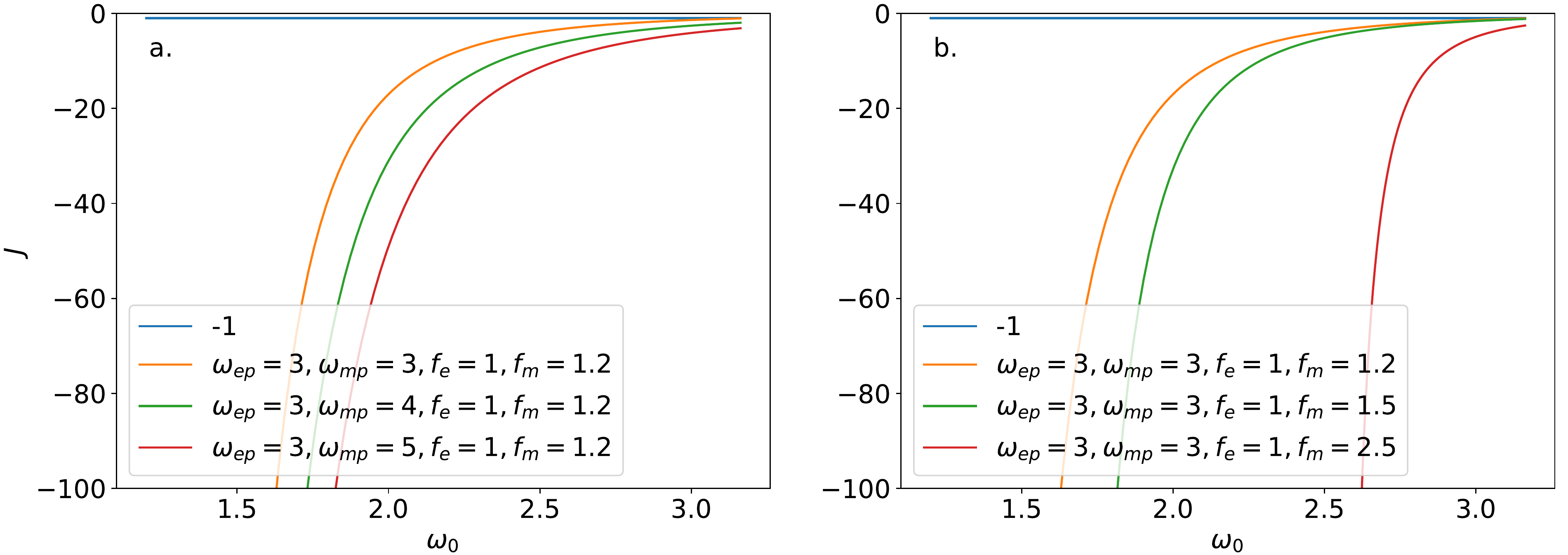}
\caption{The variation of $J(\omega_0)$ with parameters $\omega_{ep},\omega_{mp},f_e,f_m$ showing that Stockman's identity is satisfied by the lossless Lorentz model in the region of negative refraction.}\label{fInequality}
\end{figure}

To check that the Lorentzian model still satisfies \eq{1.0} we can evaluate the integral

\begin{gather}
\nonumber J(\Go_0)=\int_0^\infty \frac{\frac{ \Go_{ep}^2}{f_e}\left[\delta(\omega_1-f_e)\right]\left(1-\frac{\Go_{mp}^2}{\omega_1^2-f_m^2}\right)+\left(1-\frac{\Go_{ep}^2}{\omega_1^2-f_e^2}\right)\frac{ \Go_{mp}^2}{f_m}\left[\delta(\omega_1-f_m)\right]}{(\omega_1^2-\omega_0^2)^2}\omega_1^3d\omega_1\\
=\frac{\Go_{ep}^2f_e^2\left(1+\Go_{mp}^2/(f_m^2-f_e^2)\right)}{(f_e^2-\omega_0^2)^2}+\frac{\Go_{mp}^2f_m^2\left(1-\Go_{ep}^2/(f_m^2-f_e^2)\right)}{(f_m ^2-\omega_0^2)^2}.
\end{gather}
Fig. (\ref{fInequality}) shows the evaluation of $J(\omega_0)$ over the region of negative refraction for various parameters in the Lorentz model. We note that in all the cases demonstrated, $J(\omega_0)$ does indeed satisfy Stockman's inequality \eq{1.0} in frequency ranges over which lossless negative refraction is present.

In these examples it can be seen that $J(\omega_0)$ is a monotonically increasing function of $\omega_0$ in the frequency range of interest. If this was always the case
it would suffice to show that $J(\omega_r)\leq -1$. To explore this inequality, we consider, without any loss of generality, $\Go_{ep}\leq\Go_{mp}$ and $f_e<f_m$. This ensures that the negative refraction frequency range is $[\Go_{mp},\sqrt{f_e^2+\Go_{ep}^2}]$. We want to evaluate $J(\Go_0)$ at $\Go_0=\Go_r=\sqrt{f_e^2+\Go_{ep}^2}$ and show that it is less than $-1$. After some manipulation we have:

\begin{gather}
\nonumber J(\Go_r)=\frac{-\Go_{ep}^2\Go_{mp}^2-f_e^2(\Go_{mp}^2-\Go_{ep}^2)-f_e^2(f_m^2-f_e^2)}{\Go_{ep}^4-\Go_{ep}^2(f_m^2-f_e^2)}\\
\leq\frac{-\Go_{ep}^4-f_e^2(f_m^2-f_e^2)}{\Go_{ep}^4-\Go_{ep}^2(f_m^2-f_e^2)}=\frac{-1-f_e^2(f_m^2-f_e^2)/\Go_{ep}^4}{1-(f_m^2-f_e^2)/\Go_{ep}^2}.
\end{gather}
In the above, the numerator is clearly less than or equal to $-1$ and the denominator is less than or equal to $+1$. If we can show that the denominator is a positive quantity, we would have proven that $J(\omega_r)$ is less than -1. We, therefore, need to show that $(f_m^2-f_e^2)/\Go_{ep}^2<1$. To see this, we note that since we are in the negative refraction region we have the following inequality:

\begin{gather}
\frac{\Go_{ep}^2}{f_m^2-f_e^2}>\frac{\Go_{ep}^2}{\Go_r^2-f_e^2}>1\rightarrow \frac{f_m^2-f_e^2}{\Go_{ep}^2}<1,
\end{gather}
thus proving that $J(\omega_r)$ is less than -1. Note that $J(\omega_r)=-1$ is achieved if $\Go_{ep}=\Go_{mp}$ and $f_e=f_m$.

A further important observation was made by Kinsler and McCall \cite{kinsler2008causality}. As it is inevitably the case that the loss is non-zero at and nearby the frequency $\Go_0$ where one has negative refraction, either
$\Gve''(\Go)$ and/or $\Gm''(\Go)$ are non zero at and nearby $\Go=\Go_0$,
while $\Gve'(\Go)$ and $\Gm'(\Go)$ are both negative to ensure negative refraction. Then the integral in \eq{1.0} diverges to $-\infty$ and no useful information can be extracted from the inequality. Kinsler and McCall then proceed to obtain an inequality that is not
subject to this deficiency, but it only provides a constraint on the global loss.

Some other points in Stockman's paper deserve attention. In addressing the question
as to whether active materials can have negative refractive index and low loss, he states that ``such a resonant behavior is described by a simple
pole of the permittivity, $\Gve_r(\Go_1) \propto [\Go_1-\Go_r+i\Gg(\Go_1)]^{-1}$''. Clearly, unless $\Gg(\Go_1)$ is a linear function
of $\Go_1$, it is not a simple pole. Of further concern is that it appears that Stockman is assuming $\Gg(\Go_1)$ is real valued
as it is referred to as a frequency dependent relaxation rate. If that were true then $\Gve_r(\Go_1)$ would not even be an
analytic function of $\Go_1$ as it does not satisfy the Cauchy-Riemann equations. This is most easily seen by considering
its inverse, $\Go_1-\Go_r+i\Gg(\Go_1)$, which should also be analytic. He uses the approximation
\beq \Gve_r(\Go_1) \propto \left[\Go_1-\Go_r+i\tfrac{1}{2}(\Go_1-\Go)^2\frac{\Md^2\Gg(\Go)}{\Md\Go^2}\right]^{-1},
\eeq{1.20}
which is only valid for frequencies $\Go$ near $\Go_1$, yet under the assumption that $\Go\approx\Go_r$ 
states this implies there is an extra pole at a complex frequency
\beq \Go_1\approx \Go+2i(\Md^2\Gg(\Go)/\Md\Go^2)^{-1}, \eeq{1.25}
which is not necessarily near  $\Go$, i.e. not necessarily in the region where the
approximation \eq{1.20} is valid.
To clarify the error one may consider the function $f(z)=ze^{iz}$ that is approximately
$z+iz^2$ near $z=0$, but since $e^{iz}$ has no zeros in the complex plane $[f(z)]^{-1}$ does not have an extra pole at $z\approx i$, nor anywhere
in the upper half complex plane. 
Additionally, away from the resonance one has to take into account that $\Gve_r(\Go_1)$ needs to satisfy
\beq [\Gve_r(\Go_1)]^*=\Gve_r(-\Go_1^*), \eeq{1.30}
where the star denotes complex conjugation. This is required to ensure $\Gve_r(\Go_1)$ is the Fourier transform of a real valued integral kernel. 

Regarding whether active materials can have negative refractive index and low loss,
Nistad and Skarr \cite{nistad2008causality} point out that from an analytic perspective
there is nothing to prevent $\Gve(\Go)\approx\Gm(\Go)\approx -1$ in a frequency window in an active material. Furthermore, experiments show
that one can have negative refraction with low loss in active materials \cite{wuestner2010overcoming, xiao2010loss},
although it is not clear that an effective permittivity and an effective 
permeability are complete descriptors of these metamaterials' behavior in the frequency range of these experiments as the ratio of wavelength to unit cell size is not very large.

Also, although not widely appreciated, we remark that for active materials a lack of analyticity of $\Gve(\Go)$ in the upper half $\Go$-plane does not necessarily 
imply a loss of causality, but instead could indicate a loss of stability which 
may physically occur. 
To see this, consider a causal unstable response with integral kernel
\beqa K(t)-\Gd(t) & = & e^{-\tau t}\quad \text{for}~t\geq 0 \nonum
          & = & 0\quad \text{for}~t < 0,
          \eeqa{1.41}
where $\tau<0$ is a negative relaxation time. It gives an associated permittivity
\beq \Gve(\Go)=\int_{-\infty}^{\infty}e^{i\Go t}K(t)\,dt \eeq{1.40}
that is not analytic in the upper half plane. The integral converges when the imaginary part $\Go''$ of $\Go$ exceeds $-\tau$, giving
\beq \Gve(\Go)=1+\frac{i}{\Go+i\tau}\quad \text{when}~\Go''>-\tau.
\eeq{1.42}
One may either analytically continue
$\Gve(\Go)$ to the rest of the upper half plane resulting in an analytic function with a pole at $\Go=-i\tau$, or alternatively say $\Gve(\Go)$ is undefined in the strip
where $\Go''$ is between 0 and $-\tau$. In either case $\Gve(\Go)$ is not analytic in the upper half frequency plane. So non-analyticity does not imply non-causality as
Stockman infers. The growth of the instability \eq{1.41} will be slow when $\tau$ is small. A pole near $\Go_r$,
where the approximation \eq{1.25} might be valid when $\Go\approx\Go_r$, could similarly be associated with a 
slow growth instability. 

Additionally, it should be noted that analyticity in the upper half plane by itself does not imply causality: one has to add that $\Gve(\Go)$
has the correct behavior as $|\Go|$ goes to $\infty$, not just for real $\Go$ but also for complex $\Go$ in the upper half plane. This is best illustrated with a noncausal
$K(t)$ that is square-integrable and zero for $t<t_0<0$. 
Then $\widetilde{K}(t)=K(t+t_0)$ is causal and square-integrable, having a permittivity
$\widetilde{\Gve}(\Go)=e^{-i\Go t_0}\Gve(\Go)$ which is analytic in the upper half
frequency plane. Therefore $\Gve(\Go)$ is also analytic in the upper half frequency plane, though it does not satisfy the required behavior as $|\Go|\to\infty$.

Finally, we mention that Pendry's concept of the perfect slab lens \cite{pendry2000negative}, which spurred most of the interest in negative refraction and in particular in the drive to obtain metamaterials with $\Gve=\Gm=-1$, is itself flawed when it comes to imaging sources close to the lens: for a monochromatic point source turned on at a distance to the lens that is less than half the lens thickness the ratio of power transmitted to power emitted approaches 0 not 1 as time increases. Furthermore, the use of active materials does not remedy the problem. Additionally, cloaking due to anomalous resonance can occur and dielectric objects close to the lens are not properly imaged. For a review of these points see \cite{mcphedran2019review} and references therein.

\section*{Acknowledgment}

GWM is grateful to the National Science Foundation for support through the Research Grant DMS-1814854. AS acknowledges support through the National Science Foundation CAREER grant \#1554033. Mihai Putinar is thanked for a helpful comment. 

\section*{References}


\end{document}